\documentclass[submission, Proceedings]{SciPost}

\hypersetup{
    colorlinks,
    linkcolor={red!50!black},
    citecolor={blue!50!black},
    urlcolor={blue!80!black}
}

\usepackage[bitstream-charter]{mathdesign}
\DeclareSymbolFont{usualmathcal}{OMS}{cmsy}{m}{n}
\DeclareSymbolFontAlphabet{\mathcal}{usualmathcal}

\begin{document}

\begin{center}{\Large \textbf{Towards a study of the effects of dynamical factorization breaking at LHCb\\}}\end{center}
\begin{center}J. D. Roth\textsuperscript{1$\star$} on behalf of the LHCb Collaboration\end{center}
\begin{center}{\bf 1} University of Michigan, Ann Arbor, USA\\* jdroth@umich.edu\end{center}

\begin{center}
\today
\end{center}


\definecolor{palegray}{gray}{0.95}
\begin{center}
\colorbox{palegray}{
  \begin{tabular}{rr}
  \begin{minipage}{0.1\textwidth}
    \includegraphics[width=22mm]{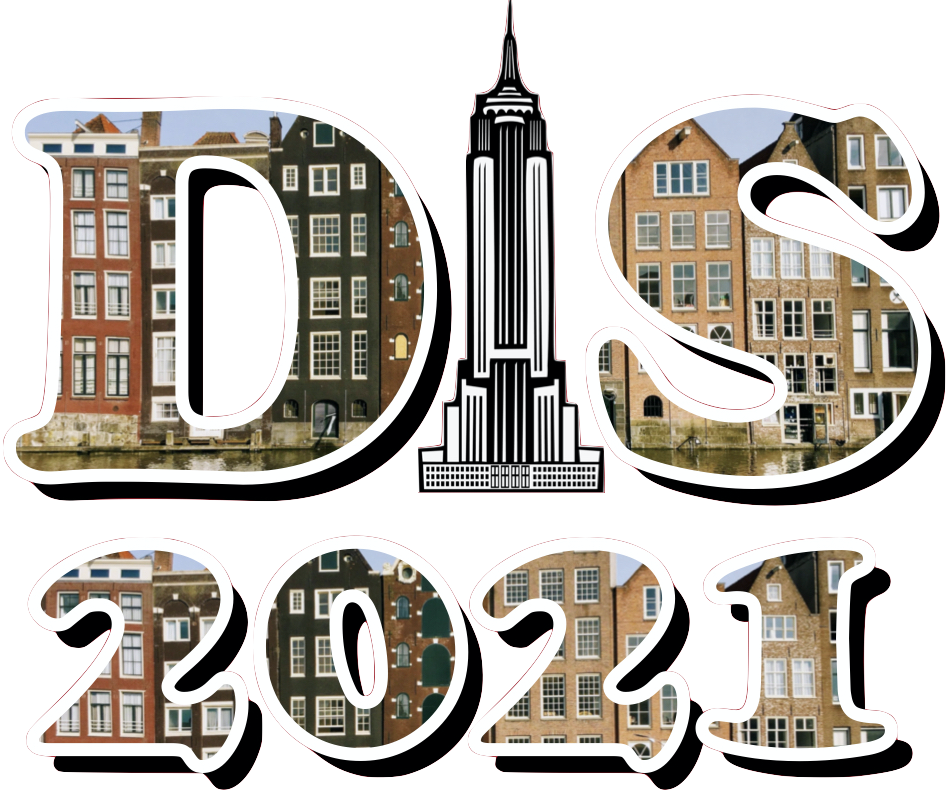}
  \end{minipage}
  &
  \begin{minipage}{0.75\textwidth}
    \begin{center}
    {\it Proceedings for the XXVIII International Workshop\\ on Deep-Inelastic Scattering and
Related Subjects,}\\
    {\it Stony Brook University, New York, USA, 12-16 April 2021} \\
    \doi{10.21468/SciPostPhysProc.?}\\
    \end{center}
  \end{minipage}
\end{tabular}
}
\end{center}

\section*{Abstract}
{\bf
    The factorization of short-distance partonic cross-sections from universal long-distance
    kinematic distributions is fundamental to phenomenology at hadron colliders. It has
    been predicted however that observables sensitive to momenta transverse to the direction
    of an energetic parton cannot be factorized in the usual way, even
    at high energies.
    It should be possible to study this factorization breaking using
    Z+jet production in high-energy proton-proton collisions by studying azimuthal correlations
    between a Z boson and associated charged hadrons. A plan to perform
    this measurement with data collected by LHCb will be discussed, along with related
    work.
}


\def\GeV{{\fam0 GeV}}%
\def\TeV{{\fam0 TeV}}%

\section{Introduction}
\label{sec:intro}
Factorization theorems are useful because they allow the computation of cross-sections for processes with non-perturbative components.
Beyond their phenomenological utility, however, factorization theorems make a formally rigorous connection between the partonic picture and the hadronic picture of Quantum Chromodynamics~(QCD).
This rigorous connection can be contrasted with the more {\it ad-hoc} connection established by the hadron formation models included in many Monte Carlo event generators.
These models are equally useful in a phenomenological sense, but they are not formally derived from QCD: they are often inspired by specific features of QCD,
but it is difficult or impossible to derive higher-order corrections to these models
or to clearly identify the kinematic regions in which they fail.
In regions where standard factorization theorems start to fail, certain techniques have already been developed to correct for effects that are typically neglected \cite{guerrero_accardi,liu_qiu,slavic_folks}.
Hadron mass corrections in particular have sometimes allowed analysts to use universal parton distribution functions and fragmentation functions to fit cross-sections that are characterized by hard scales of only a few $\GeV$
at colliders and fixed-target experiments with center-of-mass energies of a few tens of $\GeV$s.
It has also been possible to develop special-purpose factorization frameworks for certain processes, like the decay of heavy hadrons \cite{hqet},
that allow for connections to perturbation theory in specific kinematic limits.

In Section~\ref{sec:tmd} we will introduce the idea of transverse-momentum-dependent~(TMD) factorization
and provide a motivation for the study of TMD factorization breaking.
In Section~\ref{sec:plan} we detail the measurement that we plan to make,
and in Section~\ref{sec:joe} we present similar measurements that have already been made.
We present our conclusions in Section~\ref{sec:end}.

\section{TMD factorization breaking}
\label{sec:tmd}
The TMD factorization framework \cite{collins_foundations} is used to compute cross-sections that depend on the component of a hadron momentum that is transverse to the direction of one of its constituent partons,
in the limit where this transverse momentum is much smaller than the largest energy scale that is relevant to the scattering process.
This transverse momentum is often called $q_T$, and the large energy scale is called $Q$:
the TMD framework was established to compute $q_T$ spectra in the limit that $q_T\ll Q$.
In certain scattering channels, however, TMD factorization is expected to break down in certain kinematic regions, even at very high energies.
In particular: a proof \cite{rogers_mulders} has been written to show that it is not possible to use the TMD framework to factorize the cross-section for the production of a pair of hadrons from a proton-proton collision
in the kinematic region where the two final-state hadrons are produced nearly back-to-back in azimuth, which is a region where $q_T$ tends to be very small.
This breakage of factorization is also expected to apply in back-to-back dijet production and $Z$+jet production,
where a colored parton coming out of the hard process can interact with the beam remnants and absorb virtual emissions from initial-state partons.

Unlike most well-known effects that complicate the factorized picture, TMD factorization breaking is not suppressed at high energies.
Therefore, at a high-energy collider, it should be easy to isolate TMD factorization breaking effects from any other type of factorization breaking.
This breakage of TMD factorization is also interesting because it does not generalize easily:
that is, there are certain factorizable observables that look very similar to observables that break factorization.
For example: proofs have shown that TMD factorization can be used to compute $q_T$ spectra in back-to-back hadron pair production in electron-positron annihilation
and also back-to-back lepton pair production in proton-proton collisions \cite{collins_foundations}.
These processes look superficially similar to back-to-back hadron pair production in proton-proton collisions.
In these processes, it may also be possible to use TMD factorization to compute a wide variety of single-differential cross-sections, in addition to $q_T$ spectra \cite{schwartz_yan_zhu}.
It is also expected to be possible to use the collinear factorization framework to compute $q_T$ spectra for hadron pair production in proton-proton collisions in the wide-angle kinematic region where $q_T$ is similar in size to $Q$ \cite{collins_qiu},
in which case a change of only the kinematic region would distinguish a factorizable observable and a factorization-breaking observable.

Cross-sections that do not factorize often tend to share certain characteristics:
it seems to become more difficult to factorize a cross-section as the number of hadrons involved in the measurement increases, or as the observable becomes less inclusive or more differential.
But, there is not yet any set of rules that is both strict and generally applicable that can describe which observables factorize under which conditions \cite{catani_deflorian_rodrigo}.
For now, a new proof must be written for more or less each observable that needs to be factorized: factorization is handled on a case-by-case basis.
Because back-to-back production of a hadron pair in proton-proton collisions breaks factorization and is also similar to processes that do not break factorization,
it might be used to bring attention to specific criteria that prohibit factorization.

\section{Plan for measurement at LHCb}
\label{sec:plan}
We plan to measure a differential cross-section for unidentified charged hadrons produced in association with a $Z$ boson and a jet.
The same reasons that motivate the study of dihadron production also motivate the study of $Z$+hadron production, which is also expected to break TMD factorization.
It is also experimentally easier to extract clean $Z$-hadron correlations because it is easy to reconstruct a $Z$ boson via its decay to $\mu^+\mu^-$:
LHCb in particular has a proven ability to precisely measure $Z$+jet cross sections \cite{lhcbzjet1,lhcbzjet2}, and the spectra of hadrons associated to a $Z$+jet pair \cite{lhcbzjetjoe}.
If the measurement is precise enough then a $Z$+hadron measurement can be compared to a dihadron measurement to test if the number of colored partons coming out of the hard vertex has an effect on the size of the factorization breaking.

\subsection{Data sample and detector description}
\label{sec:plan_data}
To make this measurement, we will use data that was collected by LHCb during 2016 with a $p+p$ center-of-mass energy of $13~\TeV$.
The LHCb detector \cite{lhcb_detector} is a single-arm forward spectrometer covering the pseudorapidity range $2<\eta<5$. The detector includes a high-precision tracking system consisting of a silicon-strip vertex detector around the $p+p$ interaction region, a large area silicon-strip detector located upstream of a dipole magnet, and three stations of silicon-strip detectors and straw drift tubes placed downstream of the magnet. The tracking system provides for momentum measurements with relative uncertainties that vary from $0.5\%$ at low momentum to $1\%$ at $200~\GeV/c$.
We will take hadron candidates from tracks that pass through all layers of the tracking system.
The detector also has electromagnetic and hadronic calorimeters: both tracks and calorimeter clusters are used as input to a particle flow algorithm that is used as part of the jet reconstruction.
Muons are identified by a system composed of alternating layers of iron and multiwire proportional chambers, and $Z$ boson candidates will be reconstructed from high-mass muon pairs.
We will require both the jet and the $Z$ boson to have large transverse momenta: above about $15$ or $20~\GeV/c$.
Only one jet will be used from each event: the jet with largest transverse momentum.
Hadrons and jets associated with the $Z$ boson should generally recoil against the $Z$ boson, so an azimuthal window of width $\pi/3$ or $\pi/4$ on the away-side of the $Z$ boson will be established as a signal region: the rest of azimuth will be used for background estimation.

\def\pout{p_{\fam0 out}}%
\def\subscript#1{^{#1}}%
\def\pthadron{p_T\subscript{h^\pm}}%
\def\ptjet{p_T\subscript{\fam0 jet}}%
\def\ptzboson{p_T\subscript{Z}}%
\def\pttrig{p_T\subscript{\pi^0,\,\gamma}}%
\def\phihadron{\phi\subscript{h^\pm}}%
\def\phijet{\phi\subscript{\fam0 jet}}%
\def\phizboson{\phi\subscript{Z}}%
\def\phitrig{\phi\subscript{\pi^0,\,\gamma}}%
\def\pIIIhadron{\mathbf{p}\subscript{h^\pm}}%
\def\pIIIjet{\mathbf{p}\subscript{\fam0 jet}}%
\def\pIIIzboson{\mathbf{p}\subscript{Z}}%
\def\pIVjet{p\subscript{\fam0 jet}}%
\def\pIVzboson{p\subscript{Z}}%
\def\numhadron{N\subscript{h^\pm}}%
\def\numzboson{N\subscript{Z}}%
\def\numzjet{N\subscript{Z{\fam0 +jet}}}%

\subsection{Kinematic variables}
\label{sec:plan_vars}
The hadronic cross-section will be binned in three kinematic variables:
\begin{equation}
    \pout \equiv\pthadron\sin\left(\phihadron-\phijet\right),
\end{equation}
\begin{equation}
    Q \equiv\left(\pIVzboson+\pIVjet\right)^2,
\end{equation}
\centerline{and}
\begin{equation}
    z \equiv{\pIIIhadron\cdot\pIIIjet\over\pIIIjet\cdot\pIIIjet} ~.
\end{equation}
Here, $\pthadron$ is the transverse momentum of the charged hadron, and $\phihadron$ and $\phizboson$ are the azimuthal coordinates of the hadron and the $Z$ boson.
The four-momenta of the $Z$ boson and the jet are written as $\pIVzboson$ and $\pIVjet$,
and the three-momenta of the jet and the hadron are written as $\pIIIjet$ and $\pIIIhadron$.
In order to mitigate uncertainies that are associated with the beam luminosity and $Z$+jet reconstruction, we normalize the hadronic cross-section by the cross-section for $Z$+jet production:
\begin{equation}
    \left. {{\fam0 d}\numhadron \over {\fam0 d}\pout\,{\fam0 d}z\,{\fam0 d}Q} \middle/ {{\fam0 d}\numzjet \over {\fam0 d}Q}\right.~.
\end{equation}

The variable $\pout$ is used to probe transverse momenta generated by the parton shower and by long-range dynamics in both the initial state and final state.
To a first approximation, none of $\pout$ comes from transverse momentum generated at the hardest scales:
in the partonic center-of-mass frame, the $Z$ boson and the outgoing parton come out exactly back-to-back.
Any transverse momentum imbalance is generated from processes characterized by smaller energy scales.
Some of the imbalance comes from transverse motion of the initial-state partons inside the protons,
much of the imbalance is generated by splittings in the parton showers in the initial and final state,
and some comes from the hadron formation process where partons that are separated in azimuth exert forces on each other.
Note that $\pout$ ignores components of the transverse momentum imbalance along the axis determined by the $Z$ boson direction of motion,
which is roughly the same as the jet axis.
The imbalance along this axis should be determined mostly by collinear aspects of the parton splitting and fragmentation process:
with a focus on the off-axis component of the imbalance, we hope that we can improve our sensitivity to uniquely transverse-momentum-dependent effects.
This is important because factorization is not expected to break in the collinear framework.

The $Z$+jet mass $Q$ is a proxy for the hard scale that characterizes the scattering.
The way that the $\pout$ distribution changes with the hard scale is described by Collins-Soper-Sterman (CSS) evolution \cite{cssone,csstwo}.
CSS evolution is similar to a TMD variant of Dokshitzer-Gribov-Lipatov-Altarelli-Parisi (DGLAP) evolution \cite{dglapone,dglaptwo,dglapthree}, with the notable exception that the CSS evolution kernels have both perturbative and universal non-perturbative components while DGLAP's evolution kernels can be computed entirely in perturbation theory.
CSS evolution takes as input the set of kinematic distributions defined in the TMD framework, probed at one hard scale, and describes how those distributions look at another hard scale.
In particular: the results of CSS evolution are only valid when the technique is applied to a factorizable distribution defined in the TMD framework.
Hence: if we can test whether or not CSS evolution can correctly model the relationship between $\pout$ distributions measured at different values of the hard scale $Q$, then we can test whether or not TMD factorization holds.
The alternative is to compute $\pout$ distributions using TMD distributions that have already been extracted from fits to $e^+e^-$ and $ep$ scatterings:
but, as of now, no such fits have been extracted with very good precision.

We also want to bin the hadronic cross-section in the fragmentation variable $z$ in order to allow calculations that compare to this measurement to exclude the low-$z$ and high-$z$ regions, where calculation is difficult.
The $z$ variable will also provide a more complete or differential picture of the hadron formation process,
which might improve the power of the analysis if factorization breaks most strongly in a sub-region of the full phase space.

In addition to measuring the differential cross-section,
we plan to fit the $\pout$ distribution in each bin of $Q\otimes z$ in order to show clearly how the shapes of the distributions evolve with $Q$ in each bin of $z$.
These fits will hopefully make it easier to make a computation that determines whether or not this measurement is consistent with the predictions of CSS evolution.

\section{Prior measurements}
\label{sec:joe}
\begin{figure}[h]
\centering
\includegraphics[width=0.46\textwidth]{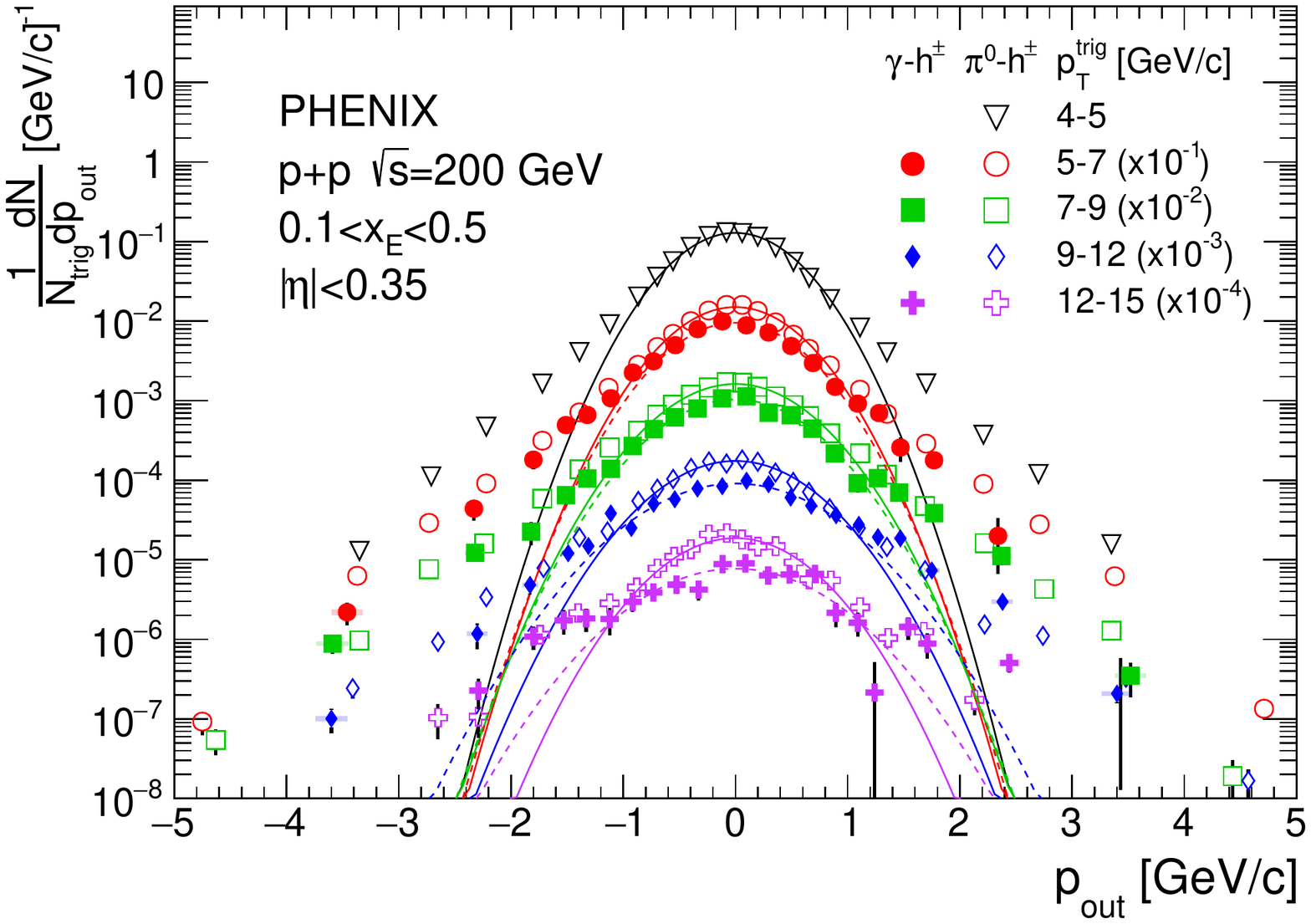}
\hfill
\includegraphics[width=0.46\textwidth]{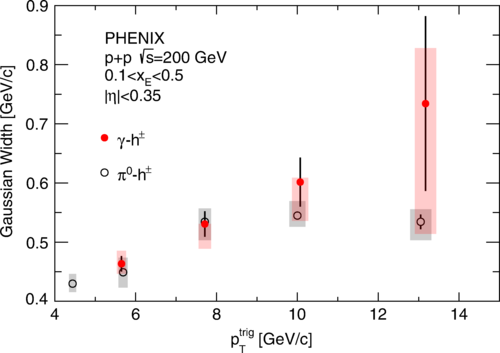}
\caption{%
    On the left are shown some of the $\pout$ distributions extracted by the PHENIX collaboration.
    The solid lines show Gaussian fits to the data points.
    On the right, the Gaussian widths are plotted against the hard scale.
    Qualitatively, the widths increase with the hard scale.
    Both plots are from Reference~\cite{joe}.
}
\label{fig:joe}
\end{figure}

Qualitatively, we have a good idea of what to expect from this measurement because similar measurements have been made at PHENIX \cite{joewrong,joe,joetwo}.
Because PHENIX uses $p+p$ collisions with lower beam energies than LHCb,
which allow less phase space for $Z$ boson production,
they measured $\pi^0$-hadron correlations and direct photon-hadron correlations instead of $Z$-hadron correlations.
Those measurements also used a slightly different set of kinematic variables, since jet reconstruction was not feasible at PHENIX
due to a limited acceptance.
In order to characterize the hard scale of the scatterings, PHENIX used the transverse momentum of the $\pi^0$ or photon instead of an invariant mass-type variable like $Q$, and
in order to estimate a fragmentation variable like $z$ they used $x_E\equiv\pthadron/\pttrig\cdot\cos\left(\phitrig-\phihadron\right)$.
Some of their results are shown in Figure~\ref{fig:joe}.
The cores of their $\pout$ distributions can be fit to Gaussian shapes, but further out in their wings the distributions fall off too slowly to fit an exponential:
this is consistent with expectations from perturbation theory for the power-law fall-off of energetic gluon radiation densities.
Qualitatively, the fitted Gaussian widths of the $\pout$ distributions increased with the hard scale.
This result matches the predictions of CSS evolution: as more energy becomes available to the particles involved in the scattering, more transverse momentum is generated in the parton showers.
No calculation has yet been published that compares the PHENIX measurement to the quantitative predictions of CSS evolution.

\section{Conclusion}
\label{sec:end}
We want to quantify the breakdown of transverse-momentum-dependent factorization, which is a fundamental prediction of QCD that has not yet been verified.
We hope that investigation into the regions where factorization fails might inspire techniques that extend a type of factorization to observables that cannot currently be factorized.
In addition, these investigations might help to develop a more general set of rules to determine which processes do and do not factorize.
Measurements that might be sensitive to the breakdown of factorization have already been made by the PHENIX collaboration,
and we plan to make another set of measurements with data from LHCb.
These measurements allow us to test for the breakdown of factorization via the breakdown of CSS evolution.
We hope that the increased availability of measurements from a variety of energy and rapidity ranges will encourage the calculation of a quantitative comparison between these measurements and the predictions of CSS evolution,
especially with the improved array of kinematic variables that the LHCb measurement will use to parameterize the scattering and fragmentation process.



\paragraph{Funding information}
This material is based upon work supported by the National Science Foundation under Grant No. 2012926.

\nolinenumbers

\end{document}